\documentclass[journal]{rmaa}

\usepackage{paralist}

\usepackage{psfrag,color,psfig}

\SetYear{}
\SetConfTitle{}

\title{Imaging and Spectroscopy of Arp 104: A Post-starburst Interacting Pair with Cross-Fuelling?}

\author{
  Nathan Roche,\altaffilmark{1}}

\altaffiltext{1}{OAN, Instituto de Astronom\'\i{}a, UNAM, Ensenada, M\'exico.}

\shortauthor{Roche N.D.}
\shorttitle{Imaging and Spectroscopy of Arp 104}

\resumen{
\item Nathan Roche: Observatorio Astronomico Nacional,
 Instituto de Astronom\'ia, Universidad Nacional
  Aut\'onoma de M\'exico, Ensenada, Apartado Postal 3--72, 22860
  Ensenada, Baja California, M\'exico (\email{ndr@astrosen.unam.mx})
\item now at: Astronomy and Cosmology Research Unit, School of mathematical Sciences,
University of KwaZulu-Natal, Durban 4041, South Africa (\email {roche@ukzn.ac.za}).
}

\listofauthors{N.D. Roche}
\indexauthor{Roche, N.D.}

\abstract{We perform $UBR$ imaging and optical spectroscopy
of the interacting galaxy pair Arp 104, at $z=0.0098$. This consists of NGC5218, a disturbed
spiral with $r_{exp}\simeq 2.53$ kpc, a substantial bulge component and an inclined outer `shell',
 the round
 spheroidal NGC5216, a connecting bridge of length $\sim 50$ kpc and a curved plume.
 Neither galaxy shows emission lines. NGC5218 has strong Balmer lines and appears to have undergone a major
 starburst ($3.2\times 10^9\rm~M_{\odot}$) some
  $\sim 0.2$ Gyr ago, triggered by the last close passage of the two galaxies. The galaxy
 is very red in its centre,
 suggesting it is dusty, but its outer regions, and the bridge connecting the two galaxies, have the blue
 colours of 0.2--0.4 Gyr old stars. NGC5216 lacks strong Balmer lines but outside its centre is blue in
 $U-B$, suggesting it experienced a star-formation episode only $\sim 40$ Myr ago.
  This could have been fuelled by gas from
  NGC5218, transfered through the bridge. The bridge passes through NGC5216 to emerge as a plume
  extending $\sim 14$ kpc to the SW of NGC5216. The plume, from its colours, is very young
  and may be a site of ongoing star-formation or the formation of a tidal dwarf galaxy.}

\addkeyword{Galaxies: Interacting}

\begin{document}
\maketitle

\section{Introduction}
\label{sec:intro}
Mergers and interactions are a very major influence on galaxy evolution,
triggering bursts of star-formation and nuclear activity, and causing permanent
morphological transformation. The evolution induced by galaxy
interactions can reveal much about the structure of galaxies and the factors
influencing their star-formation rate (SFR).
 For example, the long-tailed pair of spirals known as `The Mice' (NGC4676) contains a large number
of star clusters formed 1.5--$2.0\times 10^8$ yr ago (de Grijs et al. 2003), an age
corresponding to the point when,
according to models of the system's kinematics,
 the two galaxies last underwent closest passage. The starburst in the Mice appears to have
  been sharply peaked in time but spatially distributed throughout the system, and
Barnes (2004) concluded that it was better represented by a model where the
 star-formation in the burst is
`shock induced' (dependent on the rate of energy input from shocks),
than by a model where the SFR is solely a function of local gas density (i.e.
the simple Schmidt law).

In this paper we present observations of the relatively little-studied
interacting galaxy pair Arp 104, consisting of a spiral (NGC5218: type SBb-pec) and a
 spheroidal (NGC5216: type E0).
The system  was first described by Keenan (1935) who
noted that the galaxies were connected by `a faint but definite band of
nebulosity', which continues beyond NGC5216 as a `short curved arm' extending
SW. It was
included in the catalog of Arp (1996) in the category of `E and e-like
galaxies connected to spirals'.

NGC5218 is at RA $13^h 32^m 10^s.4$, Dec $+62:46:04$ and NGC5216 lies 4.05
arcmin away at RA $13^h 32^m 06^s.9$, Dec +62:42:02. The
respective redshifts (recession velocities) from the Sloan Digital Sky
Survey (2003) are  $z=0.009783$ ($2933\pm 22$ km $\rm s^{-1}$)
 and $z=0.009804$ ($2939\pm 26$  km $\rm s^{-1}$); these are almost identical, meaning the relative motion
of the two galaxies is approximately transverse to the line of sight with
very little radial component. For $H_0=70$ km $\rm s^{-1}$
the proper distance is 41.9 Mpc, the distance modulus
33.13 mag, and 1 arcsec will subtend 201 pc.

NGC5218 was detected by IRAS with $F_{60\mu \rm m}=6.908\pm 0.276$ Jy and
$F_{100\mu \rm m}=14.11\pm 0.56$ Jy
(Moshir et al. 1990). From the relation of Helou et al. (1985), the IRAS fluxes correspond to
 a total far-infra-red flux $F_{FIR}\simeq 1.26\times 10^{-11}(2.58f_{60}+f_{100})=4.02\times
10^{-10}$ ergs $\rm cm^{-2}s^{-1}$ and from this the luminosity
$L_{FIR}=8.59\times 10^{43}$ ergs $\rm s^{-1}$. From the relation of Kennicutt (1998)
this corrresponds to a SFR of
$3.9 \rm ~M_{\odot}yr^{-1}$. NGC5216 was not detected in the FIR.

Arp 104 was imaged in HI with the VLA as part of the `rogues gallery'
of Hibbard et al. (2001). The 21cm map of the system
 shows the HI distribution to be
 strongly peaked within the disk of
NGC5218, with
substantial amounts of gas also associated with the faint northern
plume of this galaxy and the bridge between the two galaxies. The HI density falls to a minimum at the
 position of NGC5216, but on the other side of this galaxy there is an increase in HI density
 associated with the curved plume
 extending to the SW.

Cullen, Alexander and Clemens (2003) observed Arp 104 in neutral hydrogen and
estimated that NGC5218 contains  $7.8\times 10^9\rm M_{\odot}$ of HI.
The HI kinematics show the sense of rotation as receding on the east side,
with a rotation curve flattening in the outer galaxy at a maximum line-of-sight velocity component of
140--150 km $\rm s^{-1}$.

Olsson, Aalto and H\"{u}ttemeister (2005)
 observed the centre of NGC 5218 in
CO 1-0. They  found  a rotating ring of molecular gas, 3.5 kpc across, with a
position angle 78 deg
and estimated its mass as $2.4\times 10^9 \rm M_{\odot}$. They concluded that
the prominent bar of this spiral, which has a similar position angle,
acts to transport gas inwards to this ring. From the high $\rm HCN/HCO^+$
emission-line ratio at the very centre they infer the possible presence of an AGN.

Cullen, Alexander and Clemens (2006) observed Arp 104 in CO 1-0 and estimated that NGC5218 contains a total
of $6.9\times 10^9 \rm M_{\odot}$ of $H_2$, whereas for NGC5216 they do not detect molecular gas
 and set an upper  limit $M_{H_2}<7.6\times 10^7\rm M_{\odot}$. They also estimate the time since the
pair  experienced a close passage (perigalacticon) as $3\times 10^8$ yr.

In this paper we aim to obtain a clearer picture of the nature and history of
this interacting system through direct imaging in $UBR$ passbands and
spectroscopy of both galaxies. In Section 2 we describe our observations and
data reduction. Sections 3, 4 and 5 will deal with the
structure, spectra and colours of the galaxies. In Section 6 we summarise, and discuss the
possible evolution of the system.
Throughout $H_0=70$ km $s^{-1}Mpc^{-1}$.
\section{Observations}
\label{sec:obs}
\subsection{Data}
All observations were performed using the 0.84m telescope at the Observatorio
Astronomico Nacional,
 situated at 2790m in the Sierra San Pedro Martir,
Baja California, Mexico. The direct imaging observations were
 obtained on the nights of
1 and 2 May 2005 using a Marconi CCD, binned $2\times2$ to give an
image area of  $1024^2$ pixels of size 0.475 arcsec. The 8.1 arcmin
 FOV is just
sufficient to accomodate both galaxies and the associated tidal features within a
single pointing.

We observed with 3 of the the standard Johnson
filters installed in the `Mexman' filter wheel of this telescope,
$U3$ ($\lambda_{mean}=3640\rm \AA$), $B3$ ($4330\rm \AA$) and
$R3$ ($6470\rm \AA$). Our total usable data consisted of
$10\times 600$s exposures
in $B$, $4\times 600$s in $R$, and one 900s plus $4\times 1200$s in $U$.
 For flux calibration, the
Landolt (1992) standard star 104-461 was observed.
We calibrate in Vega-system
 magnitudes and use these throughout this paper. Using the filter
response curves we calculate the
conversions into the AB system as  $(U,B,R)_{AB}=
(U,B,R)_{Vega}+(0.765,-0.062,0.181)$.

The spectroscopic observations were carried out on the night of 20 May 2005 using a
small Bollers and Chivens spectrograph (known as
`Bolitas'), fitted with a SITe1 CCD ($1024^2$  pixels, used
unbinned) and an RGL grating set at an angle of approximately 6 degrees
to cover the spectral range
$3500\rm \AA$ to $5860\rm \AA$.
 A slit of width $160\rm \mu m$ was used, aligned E-W.
 To obtain a relative
flux calibration we obtained (through the same slit) spectra of the
 spectrophotometric standard star
 Theta Virginis. For wavelength calibration, spectra were taken of a
HeAr arc lamp.
 In total, we obtained $6\times 1200$s exposures with
 the slit on NGC5218 and $4\times 1200$s exposures on NGC5216.

\subsection{Data Reduction: Imaging}
\label{sec:reduc2}
Data reduction was carried out using IRAF. The direct imaging
data were debiased, and then flat-fielded using twilight sky flats. The 4 $R$-band
and 4 $U$-band exposures, all taken on the clear night of 2 May, were
spatially
registered and combined with `sigclip' cosmic-ray rejection. Of our 10 $B$-band
exposures, 3 had been obtained on 2 May but the other 7 on the night of 1 May, when thin cloud was
present, resulting in somewhat
non-photometric conditions. We attempt to correct our photometry for this by comparing
 the fluxes of bright (non-saturated) stars in each of these seven
images with their mean fluxes on the 3 $B$-band exposures from 2 May. Following
this, the $B$-band exposures from 1 May were scaled up, and weighted down, by compensatory
factors of between 1.059 and 1.225, and then combined with the 2 May data. This produces a combined
$B$-band image with a normalization closely approximating that of the 2 May data, or that
which would have been obtained in photometric conditions.

The resolution of our combined images is approximately  1.65 arcsec FWHM
for $B$ and 1.9 arcsec in $R$ and $U$.

From imaging of the standard star 104-461 on 2 May we derived photometric
zero-points, and applied to these a correction for Galactic extinction at this position
in the sky; 0.113 mag in $U$, 0.089 mag in $B$ and 0.055 mag in $R$, according
to Caltech NED. This gave  corrected zero-points, for 1 count $\rm sec^{-1}$,
of 20.638 in $U$, 22.438 in $B$ and 22.235 in $R$.

\subsection{Data Reduction: Spectra}
\label{sec:reduc2}
The spectroscopic data were debiased and  flat-fielded.
Twilight sky flats were taken through the same slit/grating
configuration as used for the observations.
In the flat-fielding of spectroscopic data the aim is to
 remove the pixel-to-pixel sensitivity
 variations but not the variation along the
wavelength direction. To do this, each pixel
 of the sky-flat was divided by the average value in its column
 (i.e. of all pixels at the
 same wavelength, averaged along the slit length), to produce a normalized
flat-field with all wavelength dependence removed.

Spectroscopic observations of the standard star Theta Virginis were
spatially registered
and combined, the 1D spectrum extracted (using IRAF
`apall'), wavelength-calibrated
(using the associated HeAr arc), and, together with the tabulated spectral energy distribution for
this star (Hamuy et al. 1994), used
 to derive a
sensitivity (relative flux calibration) function.

The 6 1200s exposures of NGC5218 and 4 of NGC5216 were spatially
registered and combined with `sigclip' cosmic-ray rejection.
The 1D spectra of the two
galaxies were extracted using `apall' (with summation across
large apertures spanning the visible widths of the galaxies),
wavelength calibrated (using HeAr arcs) and flux calibrated in
relative $F_{\lambda}$ (using Theta Virginis).
 For each spectrum, we estimated an error function  by extracting
spectra from the individual exposures, finding the scatter between these at each pixel,
 dividing by $(N_{exp})^{1\over 2}$
and (as this was noisy) smoothing in wavelength.
\section{Structure of NGC5216 and NGC5218}
\label{sec:struc}
Figure 1 shows our combined 6000s $B$-band image of the Arp 104 system.
Structurally, NGC5216 appears to be a simple
round spheroidal, whereas the spiral
NGC5218 is much more complex with a nucleus, disk, bar, and an outer `shell'.
The disk appears to lack distinct spiral arms but instead shows a ring of diameter $\sim 68$  arcsec,
about the length of the bar.
The shell is an ususual feature for a spiral and is greatly (35--40 degrees)
inclined with respect to the disk plane, implying it originated in a major perturbation,
 presumably  an earlier  close encounter with NGC5216. A further conspicuous interaction
 feature is the bridge connecting the two
galaxies, which appears to pass through NGC5216 and extend SW to form a bright plume,
curving increasingly towards its tip some 70 arcsec from the spheroid's centre.
In the other direction, the bridge continues beyond NGC5218 northward as a fainter, more diffuse plume.
\begin{figure}
\psfig{file=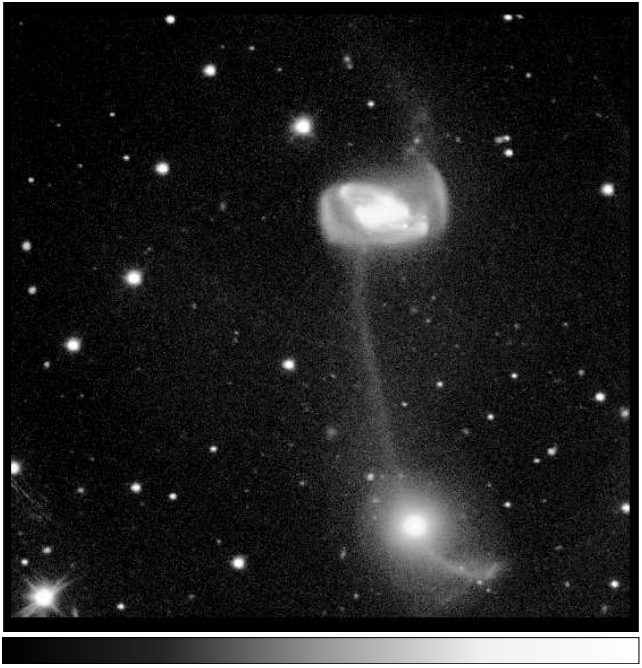,width=85mm}
\caption{$B$-band image of Arp 104 with
total 6000s exposure time. The intensity scale is logarithmic.
The image  shows a $7.89\times  7.73$ arcmin area with
 NGC5218 (upper) and NGC5216 (lower). N is top, E is left.}
\end{figure}
\subsection{Isophotes}
\label{subsec:iso}
As a first characterization of the structure of the galaxies, we fitted the $B$-band images
with concentric elliptical isophotes, using IRAF `isophot.ellipse'. Figure 2 shows
for NGC5218
the isophote
surface brightness (SB), ellipticity and position angle (PA) as a
function of semi-major axis ($r_{sm}$), together with
the best-fitting
exponential profile, which has a scalelength 14.68 arcsec (2.95 kpc) and central
SB 20.19 $B$ mag $\rm arcsec^{-2}$.

 However, the profile of NGC5218
 does not
closely follow a single exponential. At $r_{sm}<16$ arcsec, the bar, with $\rm PA\simeq 115^{\circ}$,
 is dominant, giving
isophotes with a high ellipticity (0.8) and a SB almost flat with radius.
At larger radii the isophotes trace an exponential-profile
 disk with ellipticity $\sim 0.47$
(corresponding to a disk inclination $\sim 30^{\circ}$), and $\rm PA\simeq 67^{\circ}$.
At the largest radii,
$45< r_{sm} <75$ arcsec, the isophotes trace the outer shell, which
is rounder with ellipticity $\sim 0.3$ and a PA
$\sim 105^{\circ}$, twisted $\sim 38$ deg relative to the disk plane.
\begin{figure}
\psfig{file=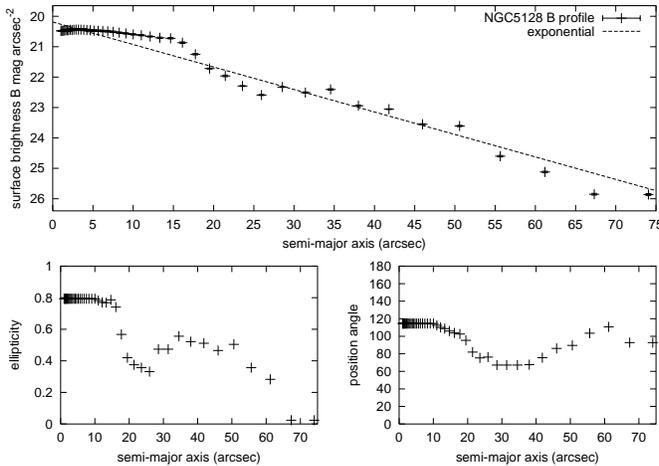,width=90mm,angle=-90}
\caption{Parameters of elliptical isophotes fit to NGC5218 in the $B$-band:
 surface brightness
against semi-major axis (points) and best-fitting
 exponential (straight line); ellipticity, and position angle (in degrees
anticlockwise from the  North).}
\end{figure}

As NGC5216 is a spheroidal rather than a disk,
 we fit the isophotal profile with a de Vaucouleurs (or Sersic $n=4$)
profile.  Figure 3 shows the
isophote SB, ellipticity and PA against $r_{sm}^{0.25}$. This galaxy is
almost round (ellipticity $\sim 0.1$), and the SB is close to the $r_{sm}^{0.25}$ law
(i.e. a straight line on this graph) out to $r\simeq 20$--25
 arcsec.
At larger radii the SB shows an excess above the $r_{sm}^{0.25}$ law, the
ellipticity rises sharply and the PA moves to $\sim 35^{\circ}$, all due to the
influence of the
bridge/plume on the isophotes at large radii.
To minimise this, we fit our model profile to only the  $0<r_{sm}<20$ arcsec isophotes.

The de Vaucouleurs profile is customarily
parameterized in terms of an `effective intensity' $I_{eff}$,
  8.325 magnitudes fainter than the SB at $r=0$, and the radius $r_{eff}$ at which the
isophote SB is $I_{eff}$. For our profile  best-fit to NGC5216,
$I_{eff}=23.70$ $B$ mag $\rm
arcsec^{-2}$ and $r_{eff}=26.2$ arcsec or 5.27 kpc.
\begin{figure}
\psfig{file=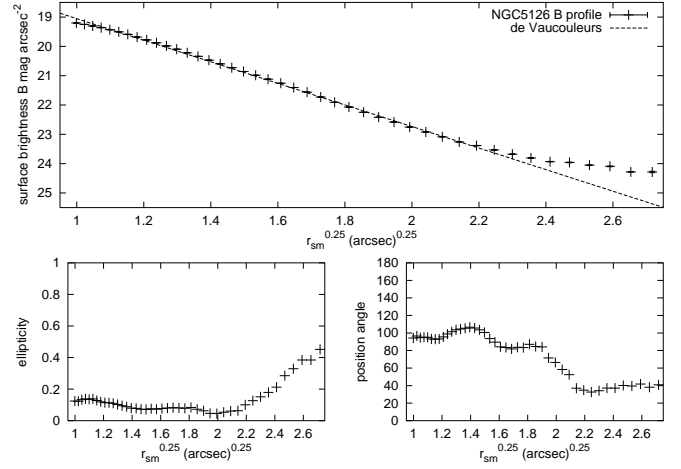,width=90mm,angle=-90}
\caption{Parameters of elliptical isophotes fit to NGC5216 in the $B$-band: surface
 brightness against $r_{sm}^{0.25}$ in $\rm arcsec^{0.25}$ (points),
 best-fitting
 de Vaucouleurs profile (straight line); ellipticity, and position angle (degrees
anticlockwise from the North).}
\end{figure}
 For NGC5218, integrating the flux out to the $r_{sm}=74$ arcsec isophote
 gives the total magnitude as $B=12.946$ with integrated
colours $U-B=0.24$ and
 $B-R=1.25$, and a half-light radius of 22.9 arcsec (4.60 kpc).
For NGC5216, integrating out to $r_{sm}=41$ arcsec gives
 $B=13.884$ with integrated colours $U-B=0.23$ (similar) and $B-R=1.44$ (redder),
 and $r_{hl}=12.0$ arcsec
(2.4 kpc).
At the distance of this system these $B$ magnitudes
 correspond to absolute magnitudes
$M_B=-20.20$ for NGC5218 and $M_B=-19.30$ for NGC5216,
 taking into account k-corrections of
0.020 mag and 0.054 mag respectively, as
calculated directly from our observed spectra
 (Section 4).

\subsection{Bulge/Disk Decomposition}
\label{subsec:bdc}
We can perform a  more sophisticated fitting
using BUDDA (Bulge/Disk Decomposition Analysis), a
program developed by de Souza, Gadotti and dos Anjos (2004). This  program fits galaxies in 2D, rather
than as a 1D radial profile,
iteratively converging on the
best-fitting combination of an exponential
disk and a bulge of generalized Sersic index. BUDDA assigns to each component a separate
flux, position angle and ellipticity. It also generates an image of the model
which can be subtracted from the real galaxy to highlight the residuals
 (e.g. bars, spiral arms, or point sources).
Using BUDDA we fit the $B$-band image of NGC5218 with a disk
plus bulge combination. NGC5216 was found to lack any significant
 disk component and so we
fitted with a single-component bulge. Table 1 gives the
parameters of the BUDDA best-fits.
\begin{table}
\begin{tabular}{lccc}
\hline
   & \multispan{2} NGC5218 & NGC5216 \\
  & disk & bulge & bulge \\
 \smallskip
Sersic $n$ & 1.0 & 1.729 & 3.679  \\
$r_{exp}$ & 12.61 & -   &  -  \\
$r_{eff}$ &  -    &  46.81 & 27.34 \\
$\mu_{cent} $ & 20.28 &  -  &  -  \\
$\mu_{eff}$ & -   &  25.28 & 23.81 \\
Ellipticity & 0.491 & 0.155 & 0.009  \\
Posn. Angle. & 68.8 & 68.7 & 86.9 \\
Lum. fraction & 0.695 & 0.305 & 1.0 \\
\hline
\end{tabular}
\end{table}

NGC5218 is a spiral with a substantial bulge component; the $B$-band
luminosity ratio is $B/D=0.429$, within and toward the upper end of the range for its
Hubble type $T=3$/Sb (e.g  Giuricin et al. 1995).
The bulge component of NGC5218
 has the same PA as the disk and is a `normal' part of the galaxy, in contrast to
the outer shell (Figure 4) for which
 the PA ($\sim 105^{\circ}$) is twisted $38^{\circ}$ anticlockwise with respect to the disk, indicating
it is produced tidally by the interaction.
For the disk component $r_{exp}=2.53$ kpc.

The BUDDA best-fit parameters for NGC5216 are very close to those obtained by fitting the isophotes with
a fixed $n=4$, and confirm the E0 classification (i.e. $n\sim4$; ellipticity$\sim 0$).
 The residuals with respect to the BUDDA model fit are a reduced
  $\chi^2$ of only 2.17 for NGC5216 but
23.51 for NGC5218, reflecting its later type and
complex and disturbed morphology.

Figures 4 and 5 show the two galaxies' residuals, i.e
the observed $B$-band images with the best fit
BUDDA models subtracted. In NGC5218 the brightest residual features ($\sim 21$ $B$ mag
$\rm arcsec^{-2}$) are at the two
ends of the bar, followed by the outer ring of the disk, and the `shell'
($\sim 23$ $B$ mag
$\rm arcsec^{-2}$) where seen well out of the disk plane.
It is also notable that the connecting
bridge curves to apparently pass through the nucleus perpendicular to the disk
and emerge as the more diffuse northern plume.

In NGC5216 the only residual associated with
the galaxy itself is a central bright spot. However, the bridge is clearly seen
to pass through the nucleus of the galaxy and emerge to the SW as a
curved plume of high SB (reaching
23.7 $B$ mag
$\rm arcsec^{-2}$), which abruptly
bends and terminates 70 arcsec (14kpc) from the nucleus.
\begin{figure}
\psfig{file=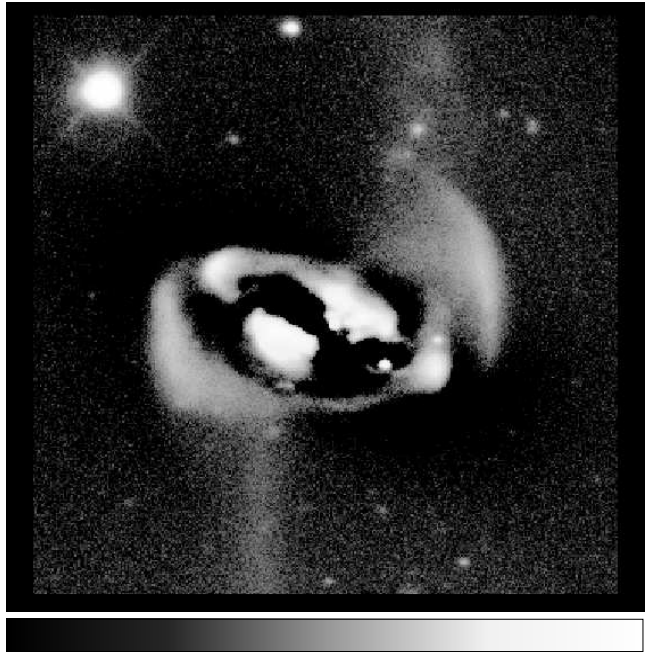,width=85mm}
\caption{$B$-band image of NGC5218 with the BUDDA best-fit disk plus bulge
  model subtracted. The area shown is $161\times 161$ arcsec,
 with a log intensity scale.}
\end{figure}
\begin{figure}
\psfig{file=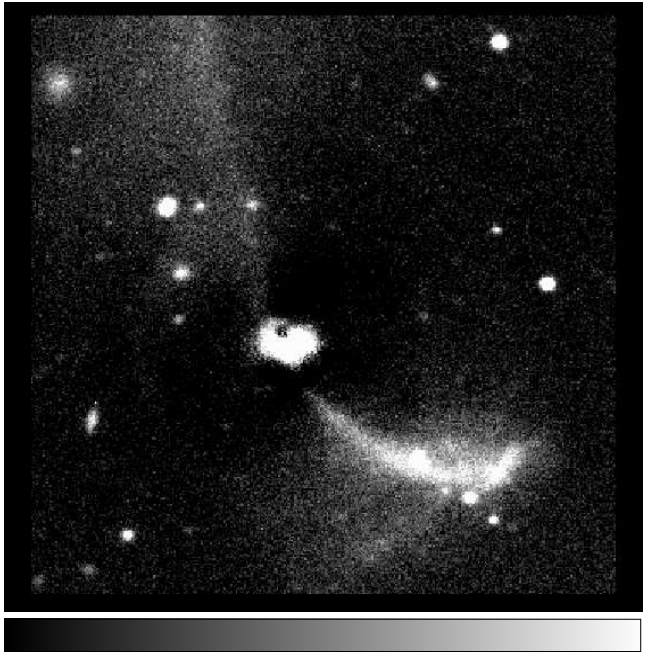,width=85mm}
\caption{$B$-band image of NGC5216 with the BUDDA best-fit bulge model
  subtracted. The area shown is  $161\times 161$ arcsec,
 with a linear intensity scale.}
\end{figure}
\section{Spectroscopy; Age-dating the starburst}
\label{sec:spec}

Figures 6 and 7 show our `Bolitas' spectra of NGC5218 and NGC5216, with detectable
features labelled. The continuum signal/noise at
 $\lambda>4200\rm \AA$
 is $\sim 18\sigma$ for NGC5218 and $\sim 9\sigma$ for NGC5216. These spectra are plotted
with  only a relative flux calibration as the slit will capture only a fraction of the flux
from each galaxy.
 However, if we assume that the region of each galaxy covered by the slit is representative
  of the whole object,
 we can estimate an approximate whole-galaxy absolute calibration by integrating each
 spectrum over the
 $B$-band response function and equating the mean value to the total $B$
 magnitude (Section 3). This gives one flux unit
 on Figure 6 corresponding to a flux from the whole galaxy of
 $0.95\times 10^{-14}$ ergs $\rm cm^{-2} s^{-1} \AA^{-1}$ or
 luminosity $2.05\times 10^{39}$ ergs $\rm s^{-1} \AA^{-1}$, and  on Figure 7, one unit is
 $1.20\times 10^{-14}$ ergs $\rm cm^{-2} s^{-1} \AA^{-1}$ or luminosity
  $2.59\times 10^{39}$ ergs $\rm s^{-1} \AA^{-1}$.
\begin{figure}
\psfig{file=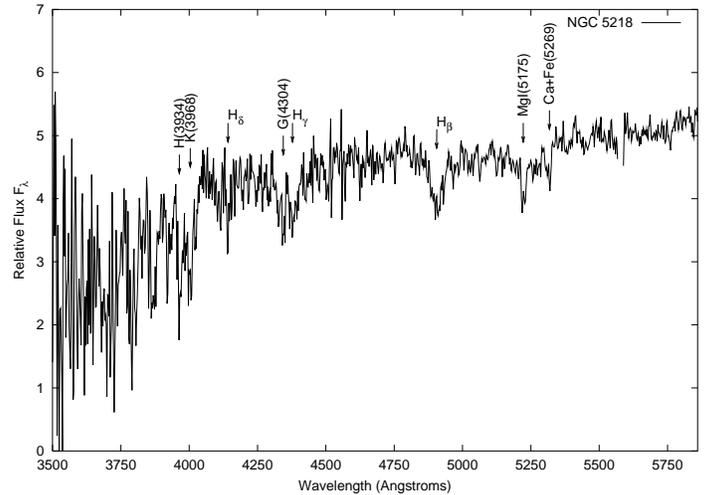,width=95mm,angle=-90}
\caption{Observed spectrum of NGC5218, from 7200s of Bolitas data (6 exposures),
 with detected lines labelled.}
\end{figure}
\begin{figure}
\psfig{file=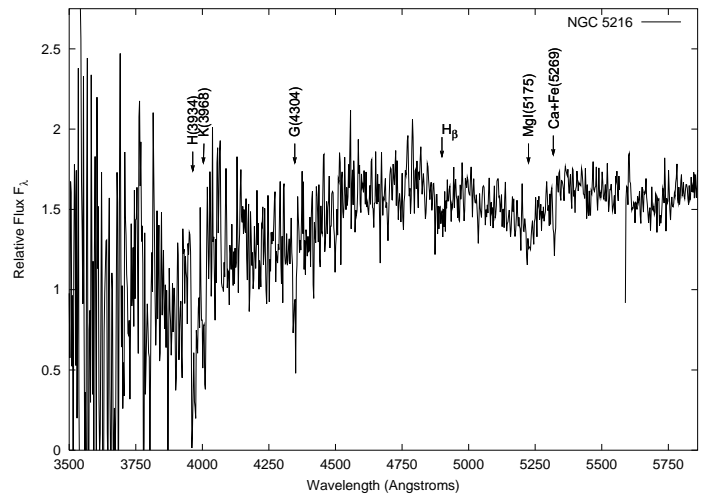,width=95mm,angle=-90}
\caption{Observed spectrum of NGC5216, from 4800s of Bolitas data (4 exposures),
 with detected lines labelled.}
\end{figure}

Table 2 gives line equivalent widths, estimated
using IRAF 'splot', for both spectra.
 The statistical significance of each line is
  estimated by summing in quadrature  our inter-exposure error function over
the FWHM of the line and dividing this by the integrated line profile.
For both galaxies, we do not detect any emission lines for either galaxy, but do find strong absorption lines.
 The most
obvious difference between the two spectra is that NGC5218 has much stronger Balmer
absorption lines:
$H\delta$, $H\gamma$ and $H\beta$, of which the last two are  significantly
broadened, with approximate FWHM of 28 and 45 $\rm \AA$,
  compared to the instrumental resolution of $\rm FWHM\simeq 16$--$18\AA$.
NGC5216 has much weaker Balmer lines, with only $H\beta$ detected,
 but its other absorption lines are
stronger than in NGC5218, as expected for an earlier Hubble type.
\begin{table}
\caption{table 2}
\begin{tabular}{lcc}
\hline
Line & \multispan{2} Equivalent width $\rm \AA$ \\
     & NGC5218   & NGC5216 \\
$\rm [OII]3727$ & $<5^*$ & $<10^*$ \\
K(3934) & $5.40\pm0.70$  & $15.3\pm 2.3$ \\
H(3968) & $6.00\pm 0.73$  & $9.84\pm 1.88$ \\
$\rm H\delta$(4102) & $2.92\pm 0.60$ & $<2.4^*$ \\
G(4304) & $4.13\pm 0.90$ & $6.48\pm 0.87$ \\
$\rm H\gamma$(4340) & $4.47\pm 1.04$  & $<2.4^*$\\
$\rm H\beta$(4861) & $8.01\pm 1.04$ & $3.74\pm 1.46$ \\
$\rm [OIII]5007$ & $<1.5^*$ & $<1.6^*$ \\
Mg(5175) & $3.33\pm 0.52$ & $5.33\pm 1.06$ \\
Ca+Fe(5269) & $2.08\pm 0.85$ & $1.87\pm 0.73$ \\
\hline
\end{tabular}

$^*$ $2\sigma$ upper limits
\end{table}

In NGC5216,
the absence of emission lines indicating star-formation is not surprising in view of the lack
 of HI in the galaxy (Hibbard 2001). However, NGC5218 is gas rich and an IRAS
 detection with the FIR flux apparently corresponding to a SFR of $\rm 3.9~M_{\odot} yr^{-1}$.
 On the basis of our approximate
 `absolute calibration', our estimated upper limit
 $\rm EW([OII]3727)<5\AA$ for NGC5218 means
 $\rm F([OII]3727)<1.3\times 10^{-13}$ ergs $\rm cm^{-2}s^{-1}$ and a luminosity
 $\rm L_{[OII]}<2.8\times 10^{40}$ ergs $\rm s^{-1}$, and $\rm EW([OIII]5007)<1.5\AA$
 means $\rm F([OIII]5007)<6.4\times 10^{-14}$ ergs $\rm cm^{-2}s^{-1}$ and a luminosity
 $\rm L_{[OIII]}<1.4\times 10^{40}$ ergs $\rm s^{-1}$.
 From Kennicutt (1998), $\rm SFR\simeq 1.4\times 10^{-41}L_{[OII]}$, so in the absence of dust extinction,
 the SFR is
$<0.4\rm M_{\odot} yr^{-1}$. The [OIII] flux depends on factors such as metallicity, as well as SFR,
 but typically $\rm F([OIII]5007)\simeq 1$--$\rm 2 F(H_{\beta})\simeq 0.3$--$0.7 \rm F(H_{\alpha})$,
 and from Kennicutt (1998), $\rm SFR\simeq 7.9\times 10^{-42}L_{H_{\alpha}}$ giving the upper
 limit 0.16--$\rm 0.37~M_{\odot} yr^{-1}$. Hence our non-detection of emission lines from NGC5218,
 would require that the FIR emission is from a source other than young stars,
  i.e. an obscured AGN, and/or that star-formation
 in NGC5218 is subject to heavy dust extinction, $A_V>2.5$ mag.

We compare the spectra of NGC5218 and NGC5216 with observed local galaxy spectra
from the database of Storchi-Bergman, Calzetti and Kinney (2004), and with model spectra
(Jimenez et al. 2004) representing
 evolving stellar populations at ages from 1 Myr to 13 Gyr.
 Comparing NGC5218 with  M31 (NGC224), which is also Sb
 type (but not barred or interacting), we find that NGC5218 has
 Balmer lines about 3 times as strong (and broader),  and it is also
 somewhat bluer (more UV continuum).
 Strong/broad Balmer lines indicate the galaxy has an enhanced
content of type `A' stars, and
  imply that it underwent a major burst of star-formation between 0.09 and 1.0 Gyr ago
    (e.g. Gonz{\'a}les Delgado, Lietherer and Heckman 1999).

We fit the observed spectrum of  NGC5218
 with a two-component model, consisting of a
sum of the observed M31 spectrum (representing the pre-interaction galaxy)
 and a Jimenez et al. (2004) model for a single
 age
stellar population of solar metallicity (representing the recent starburst), both redshifted
 to $z=0.0098$. This model is parameterized in terms of $T_{sb}$, the age of the starburst,
 and $f_{sb}$, the starburst component's fraction of the total flux at
 $\lambda_{restframe}=4500\rm \AA$. We fit  by minimizing $\chi^2$, weighting each point
 using the inter-exposure errors.

The best-fit (reduced $\chi^2=1.22$) is found
for $T_{sb}=0.1^{+0.1}_{-0.04}$ Gyr and $f_{sb}=0.57\pm 0.03$.
However, this age could be an underestimate if M31 is slightly
too red a template. The broadness of the $H\beta$
line favours a greater age, and a model fit to the $H\beta$ profile in isolation
gives an age $0.4\pm 0.2$ Gyr. We therefore adopt
 the intermediate age 0.2 Gyr as our best estimate, with $f_{sb}=0.54$ (the best-fit for this age).
Figure 8 shows this model, illustrating how the sum of these two components closely
approaches the observed
spectrum.

The starburst, in order to produce $54\%$ of the present-day blue flux,
would have absolute magnitude $M_B\simeq -19.53$,
 which for the Jimenez et al (2004)
model of age 0.2 Gyr and Salpeter IMF would require a stellar mass
$3.2\times 10^9~\rm M_{\odot}$. By comparison, the total luminosity of NGC5218
is $1.8\times 10^{10}\rm L_{\odot}$, and from
the model of Bell and de Jong (2001), for a spiral with $B-R=1.25$ the stellar $M/L_B\simeq 2.2$, giving
 $\sim 4\times 10^{10}M_{\odot}$. From this we estimate that of order
 $8\%$ of the current stellar mass was formed in the burst.

 If the burst duration was
 $\sim 100$ Myr, its mean SFR would have been
  $\rm \sim 32\rm ~M_{\odot}yr^{-1}$), about an order of magnitude
 greater than either the FIR upper limit for the current SFR ($\rm 3.9~M_{\odot}yr^{-1}$),
   or the time-averaged  SFR (stellar mass/galaxy age) for the whole galaxy
  ($\rm \simeq 3$--$\rm 4~M_{\odot}yr^{-1}$). Hence HGC5218 underwent a major starburst, but
   neveretheless this
 consumed only a fraction of the total gas content of the  pre-interaction
 galaxy, as NGC5218 is still very gas rich with $M_{\rm HI+H_2}\simeq 1.47\times 10^{10}M_{\odot}$
 (Cullen, Alexander and Clemens 2003, 2006).

We fit the observed spectrum of NGC5216 with a combination of the observed spectrum of M32,
a non-interacting elliptical without recent star-formation, and the Jimenez et al (2004)
 single age population. The best-fit model (reduced $\chi^2=1.15$)
 has $T_{sb}=0.040_{-0.025}^{+0.060}$ Gyr and $f_{sb}=0.37\pm 0.06$. Ages $>0.1$ Gyr are excluded, and
  would cause
  stronger Balmer lines than observed.
 Figure 9 shows the best-fit model. Here the starburst component would have $M_B=-18.22$
 and a stellar mass $2.56\times10^8 M_{\odot}$.

\begin{figure}
\psfig{file=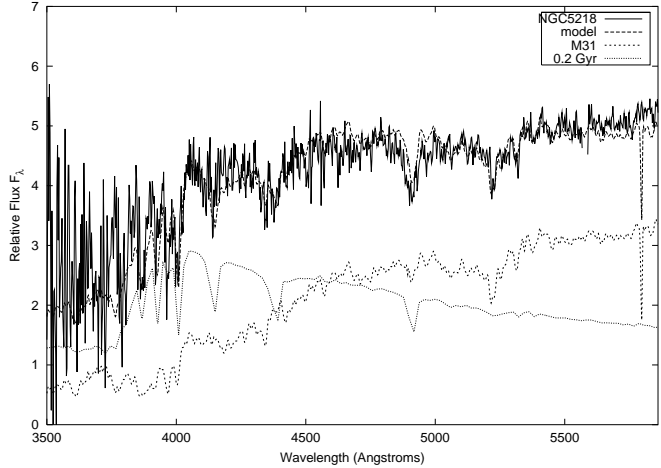,width=90mm,angle=-90}
\caption{Observed spectrum of NGC5218 (solid) together
with a best-fitting
model (heavy-dashed) which is a normalized
 sum of the observed spectrum of M31
(Storchi-Bergman et al. 2004: light-dashed) and a model spectrum
for a 0.2 Gyr age stellar
population (Jimenez et al. 2004; dotted).}
\end{figure}
\begin{figure}
\psfig{file=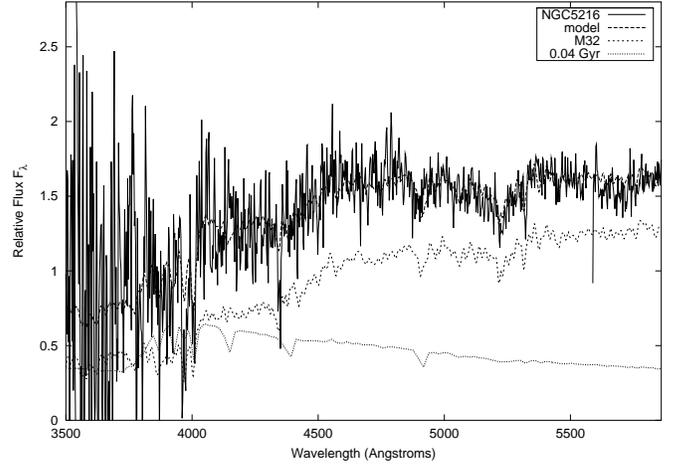,width=90mm,angle=-90}
\caption{Observed spectrum of NGC5216 (solid) together
with a best-fitting
model (heavy-dashed) which is a normalized
 sum of the observed spectrum of M32
(Storchi-Bergman et al. 2004: light-dashed) and a model spectrum
for a 0.04 Gyr age stellar
population (Jimenez et al. 2004; dotted).}
\end{figure}

\section{Colours}
\label{sec:col}
Using `isophot.ellipse' we can measure, on the $R$ and $U$
images, the mean SB on the set of isophotes previously fitted on the $B$-band image. The difference
in two passbands' SBs for each isophote  will then give its colour $U-B$ or $B-R$. Figure 10 plots isophote
colour against $r_{sm}$ for the two galaxies.
For NGC5216, we also measure the isophote SBs and colours with the bridge/plume (including the section passing
 through the galaxy)  masked out in all
passbands. This is to determine the colour of the body of the spheroidal, without `contamination' from
the interpenetrating bridge/plume, which might be very different in colour.

Both galaxies are significantly
bluer at larger radii, in both colors. The spiral NGC5218
has a much stronger $B-R$ gradient than NGC5216; compared to the spheroidal it is about 0.26 mag
 redder at its centre but is
bluer at all $r_{sm}>17$ arcsec.
 However,in $U-B$, the spheroidal NGC5216 shows the stronger
colour gradient, and outside of its central $r_{sm}\sim 3$ arcsec is
 bluer  than NGC5218.
We also see that masking out the bridge/plume passing through NGC5216
has little effect on its isophotal colours, producing only a very slight reddening. This confirms that
 the strong
$U-B$ gradient and blueness of NGC5216 are intrinsic to its own stellar content and
not due to blue flux from the plume.
\begin{figure}
\psfig{file=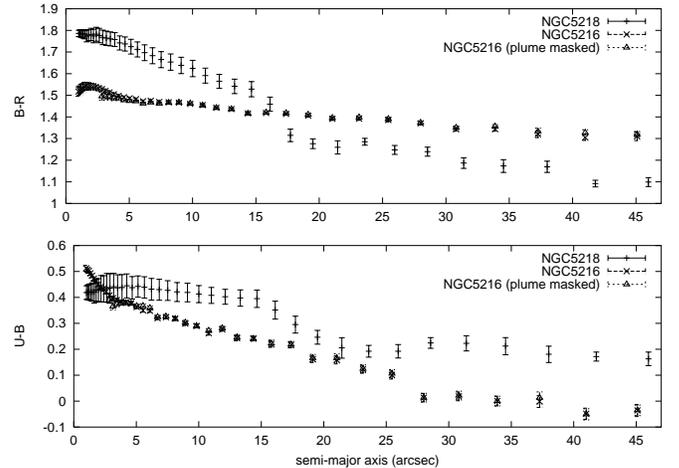,width=90mm,angle=-90}
\caption{Colours (a)$B-R$ and (b) $U-B$ of isophotes as a function of
  semi-major axis, for NGC5218, NGC5216, and NGC5216
  with the bridge/plume masked out}.
\end{figure}

Using polygon
  photometry (IRAF `polyphot') we measure $B=16.46$, $B-R=0.65$
  and $U-B=0.12$ for the central part of the bridge,
   which is much bluer than the integrated colours of either galaxy.
 For the bright plume curving SW from
  NGC5216 (excluding the part $<33$ arcsec from the
  galaxy centre,  to reduce its contamination of the colours) we measure
$B=17.53$ with $B-R=1.04$ and $U-B=-0.16$, again much
bluer than either of the galaxies. For the tip of the plume only
 (from the `bend'
to its final end) we measure $B=18.72$ with similar colours $B-R=0.86$, $U-B=-0.14$.
We note the plume is redder in $B-R$ but bluer in $U-B$ than the bridge.

We calculate $U-B$ and $B-R$ colours for the Jimenez et al. (2004)
single-age, solar metallicity  model spectra
redshifted to $z=0.0098$, for all ages 0.001 to 14 Gyr.
Figure 11 shows the locus of model colours
  together with the observed colours (integrated and
for individual isophotes) of the two galaxies and of the tidal features.

The integrated colours of NGC5216 are similar to a single 5 Gyr age stellar
 population, but also lie on a straight line between $\sim 10$ Gyr age
and very young
(0.02--0.04) Gyr stars, and hence could be produced by a mixture of these, as our fit to
the spectrum suggests. The integrated colour of NGC5218 cannot be fitted by any
 single age model, but is
 consistent with a mixture of old ($\geq 5$ Gyr) and 0.2--0.4 Gyr age stars, again in
 agreement with our interpretation of its spectrum. The bridge colours
closely match a pure 0.4 Gyr old population. However, the plume colours are
 more consistent with a mixture of very old stars (which were probably carried out of NGC5216) and a greater
 number of extremely young ($\sim 0.02$ Gyr) stars.
Considering that the plume, unlike NGC5216, corresponds to a concentration of HI, it could still be a
 site of star-forming activity.

Looking at the colours of the individual isophotes, the red centre of NGC5218 has the
colour of 5--14 Gyr old stars with dust reddening of about 0.2 in $B-R$. With increasing radius the isophote
colour gradually shifts bluewards towards that of the bridge, or a 0.2--0.4 Gyr
age population.
The central colours of
NGC5216 are consitent with very old ($>10$ Gyr) stars, but with no requirement for dust-reddening.
With increasing radius the isophotal
colour shifts rapidly bluewards in a direction indicating an increasing relative contribution
of very young (0.02--0.04 Gyr) stars. We can see from Figure 11 how the younger age of the young stars in
NGC5216, compared to NGC5218, can  account for the spheroidal's stronger $U-B$ colour gradient,
despite its weaker $B-R$ gradient. From
age 0.02 Gyr to 0.4 Gyr the models show much reddening in $U-B$, from -0.47 to +0.10, but little change in
$B-R$, from 0.58 to 0.68. We note again that the young stars are present in the body
 of NGC5216
and not just the bridge/plume.

In summary, the tidal features in the Arp 104 system are blue, as we would expect, but the
most unusual features of the colours are (i) the redness of the NGC5218 nucleus, implying dust, and (ii) the
blueness (outside the nucleus) of the spheroidal NGC5216, implying very recent star-formation.
\begin{figure}
\psfig{file=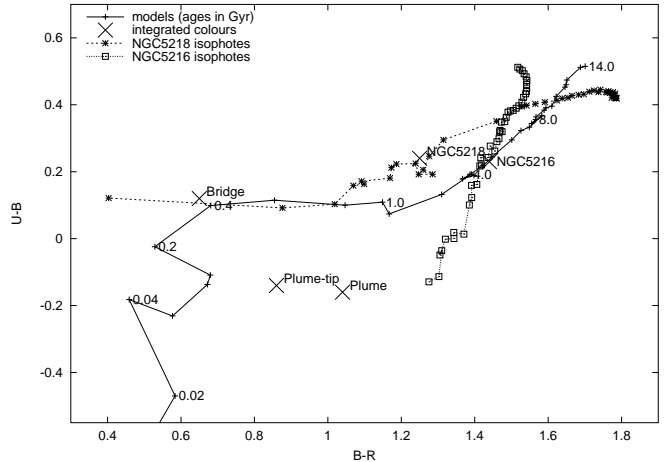,width=90mm,angle=-90}
\caption{$B-R$ against $U-B$ colours for Jimenez et al. (2004) single-age models, ages labelled in Gyr;
integrated colours of the two galaxies and some tidal features; and the loci of the isophotal
colours for the two galaxies, in both cases running from
 the galaxy centres at the top left (reddest), to the outermost edges at the bottom right (bluest)}
\end{figure}
\section{Discussion and Conclusions}
We have obtained $UBR$ imaging and optical spectroscopy
of the interacting galaxy pair Arp 104 (NGC5218 and NGC5216).
Examining the profiles we confirm that NGC5218 is a disk galaxy with a substantial bulge component
($B/D\simeq 0.43$) and that
NGC5216 is a de Vaucouleurs profile E0
For both galaxies,  we detect no  emission lines, and,
in the absence of dust extinction estimate upper limits to the SFRs of
0.2--$\rm 0.4~M_{\odot}yr^{-1}$. Cullen, Alexander and Clemens (2003) do claim a detection of $\rm H\alpha$
emission for NGC5218, but with a very low luminosity implying a SFR $\rm \sim 0.05~M_{\odot}yr^{-1}$.

However, the FIR luminosity of NGC5218 may be evidence of a SFR of up to
$\rm 3.9 ~M_{\odot}yr^{-1}$,
 and the galaxy contains a large mass of hydrogen,  $M_{\rm HI+H_2}\simeq 1.47\times 10^{10}M_{\odot}$
 (Cullen, Alexander and Clemens 2003, 2006).
Furthermore the very red colours of the inner $\sim 10$ arcsec, up to $\Delta(B-R)\simeq 0.2$ mag
 redder than  10 Gyr old
 stellar population, imply there is some dust reddening. The red region corresponds to the
 molecular gas bar described by Olsson, Aalto and H\"{u}ttemeister (2005). Hence it seems possible
  there is ongoing star-formation of anything up to  $3.9\rm ~M_{\odot}yr^{-1}$ occurring deep in
   this nucleus or bar, with heavy extinction of
   $A_V>2.5$ mag. Alternatively some or most of the FIR emission may be from
    a dust-shrouded AGN. Mid infra-red spectroscopy might distinguish the two.

    Whether or not NGC5218 is currently forming stars, it seems to have undergone a burst of much more intense
    star-formation (an order of magnitude above the time-averaged SFR)
    at an earlier stage in its interaction, probably at or shortly after the perigalacticon with NGC5216,
     an estimated 0.3 Gyr ago (Cullen, Alexander and Clemens 2006).
      Firstly, the spectrum shows strong Balmer absorption lines, especially $H\beta$, consistent with a major
      starburst 0.2--0.4 Gyr ago; from our model fit to the spectrum we estimated the mass of stars formed
      as     $3.2\times  10^{9}M_{\odot}$. Secondly, both the inclined outer shell around the disk of NGC5218,
      and the bridge connecting the two galaxies, have colours consistent with the same age, 0.2--0.4 Gyr.
      The close passage must have disrupted the disk, triggered starbursting
      within it and created the shell, and then as the two galaxies separated in their orbit,
       NGC5216 pulled away substantial amounts of gas from NGC5218 together with newly formed
       blue stars from the shell or outer disk, forming the  bridge.

      Comparing the
       spectrum of NGC5216 with the elliptical M32, it shows an excess of blue/UV flux, suggesting the presence of
       young stars, but the Balmer lines are not especially strong, favouring a younger age $<0.1$ Gyr.
        We best-fit our spectrum with a $\sim 40$ Myr age
       starburst of much lower mass ($2.56\times 10^8\rm  M_{\odot}$) than in the spiral.
         In moving the $\sim 50$ kpc  from one galaxy to the other in $\sim 0.2$ Gyr,
 the focus of star-formation has moved at $\sim 240$ km $\rm s^{-1}$,
 about the rotation velocity of the spiral.
        The $U-B$ profile of NGC5216 suggests these young stars are distributed throughout the body of the
        spheroid but are more prominent at larger radii and much
less so in the central $\sim 3$ arcsec, which is red.

 The unusual presence of young stars in this E0-type
- formed 0.2--0.3 Gyr after the close passage  -
can be explained if $\rm \sim 2\times 10^{8}M_{\odot}$
of HI was transferred from the spiral, along the bridge to NGC5216, {\it refuelling}
 the spheroidal and
resulting in a short-duration starburst.
 It is plausible that the burst in NGC5216 could have been fuelled entirely through
	the bridge, as similar bridges e.g. in NGC2992/3 (Duc et al. 2000) may contain the required HI mass in
	only $\sim 10kpc$ of length and
	evidence of spiral-to-elliptical cross-fuelling along tidal
	bridges, at rates $\rm \sim 2~M_{\odot}
	 yr^{-1}$,
	 has been reported for other interacting pairs, such as Arp 105 (Duc et al. 1997).
	NGC5216 now shows no emission lines and contains very little
	HI or $\rm H_2$, and our
	spectra and photometry support the suggestion of Cullen, Alexander and Clemens (2006)
	that cross-fuelling did occur in this system but that the transferred gas has all been exhausted.

	However, some gas, rather than forming stars within NGC5216, appears to have
	 passed through and out of the galaxy, where the bridge becomes the
	 SW. The plume is very blue in $U-B$ implying its
	 stellar content is  dominated by a
	 very young ($<40$ Myr) stellar population, perhaps  with a smaller content of old stars.
	The tip of the plume, 60--70 arcsec from the nucleus of NGC5216,
	is broadened and the HI map shows a
	 concentration of HI, resembling the HI peak at the
	 end of the NGC2992
	 tidal tail (Duc et al. 2000), which is both a star-forming region and a formative tidal dwarf galaxy.
	  Extrapolating the estimated rate of movement of the focus of star-formation,
	   from the centre of NGC5216
	 it would take $\sim 50$ Myr to reach the plume tip, which  takes us to approximately
	  the present day  and so is consistent with the
	 plume tip being  the current site of star-formation in the system. The apparent magnitude we
	  measure for the plume tip corresponds to $M_B\simeq -14.43$, which
	   for an age $\sim 0.04$ Gyr is a stellar mass of only $\sim 10^7 \rm M_{\odot}$. Hence
	    if this is a tidal dwarf galaxy, which typically
	    have virial masses $\sim 10^9 \rm M_{\odot}$ (e.g. Braine et al. 2001),
	     it is mostly gas, and in a very early stage of its star-formation.

	 To verify the existence of a formative tidal dwarf galaxy at the plume tip,
	 observing in $\rm H\alpha$ with a Fabry-Perot or another instrument of  high velocity
	 resolution (e.g. the Manchester Echelle Spectrography),
	 could reveal the spatial extent of the
	  current star-formation and whether the star-forming region shows the steep
	 ($\sim 50$ km $\rm s^{-1}$)  velocity gradient expected if the plume tip is sufficiently
	  massive ($\rm \sim 10^9~ M_{\odot}$) and self-gravitating that it is destined
	  to become an independent dwarf galaxy
	  (e.g. Bournard et al. 2004).
	  In several $10^8$ yr NGC5216 and NGC5218 will experience another close encounter,
	  perhaps this time colliding and merge, and could then eject the `plume tip object' to become
	   a separate galaxy.
\newpage
\acknowledgements
          NR acknowledges the support of the Universidad Nacional Autonoma de Mexico, and the help of all
	  at the Observatorio Astronomico Nacional, San Pedro Martir, especially Michael Richer, in the
	  observations described here.

\end{document}